\documentclass[aps,prb,twocolumn]{revtex4-1}

\usepackage{graphicx} 
\usepackage{amssymb,amsmath} 
\usepackage{verbatim} 
\usepackage{mathrsfs}

\newcommand{\brho}{\boldsymbol{\rho}}
\DeclareMathOperator{\Tr}{Tr}

\usepackage[usenames,dvipsnames]{color}
\usepackage[normalem]{ulem} 

\graphicspath{{figs/}}
\DeclareGraphicsExtensions{.pdf}

\newcommand{\sanseb}{University of the Basque Country UPV/EHU, Nano-Bio Spectroscopy Group, Avenida de Tolosa 72, 20018 San Sebastian, Spain}
\newcommand{\hamburg}{Max Planck Institute for the Structure and Dynamics of Matter, Luruper Chaussee 149, 22761 Hamburg, Germany.

Center for Free-Electron Laser Science and Department of Physics, University of Hamburg, Luruper Chaussee 149, 22761 Hamburg, Germany
}

\begin{document}

\title{A first principles TDDFT framework for spin and time-resolved ARPES in periodic systems}
\date{\today}

\author{Umberto De Giovannini}
\email{umberto.degiovannini@gmail.com}
\affiliation{\sanseb}

\author{Hannes H\"ubener}
\email{hannes.huebener@gmail.com}
\affiliation{\sanseb}

\author{Angel Rubio}
\email{angel.rubio@mpsd.de}
\affiliation{\sanseb}
\affiliation{\hamburg}

%
%
\begin{abstract}
We present a novel theoretical approach to simulate spin, time and angular-resolved photoelectron spectroscopy (ARPES)
from first principles that is applicable to surfaces, thin films, few layer systems,
and low-dimensional nanostructures.
The method is based on a general formulation in the framework of time-dependent density functional theory (TDDFT) to 
describe the real time-evolution of electrons escaping from a surface under the effect of any 
external (arbitrary) laser field. 
By extending the so called t-SURFF method to periodic systems one can calculate the final photoelectron spectrum by collecting the flux of the ionization current trough an analysing surface.
The resulting approach, that we named t-SURFFP, allows to describe a wide range of irradiation conditions without any assumption on the dynamics of the ionization process allowing for pump-probe simulations on an equal footing.
To illustrate the wide scope of applicability of the method we present applications to graphene, mono- and bi-layer WSe$_2$, and hexagonal BN under different laser configurations.

\end{abstract}
\maketitle

%
%
\section*{Introduction} \label{sec:intro}

Angular resolved photoelectron spectroscopy is one of the most prominent 
and mature techniques employed to probe the electronic properties of crystalline materials.
In its most traditional application it allows to directly map band dispersion and Fermi surfaces of 
solids from the energy and momentum distribution of the escaping electrons~\cite{Hufner:2003hj,Damascelli:2003kq}.

Following the advances in laser pulse generation and photoelectron detection techniques
in the last years we have witnessed an increasing presence of ARPES experiments capable to resolve 
time and spin polarization degrees of freedom~\cite{Krausz:2014gc,SanchezBarriga:2016hl}.
Time-resolved ARPES (tARPES) unlocks the time degree of freedom to study non-equilibrium dynamic of 
solids at the natural time scale of electronic excitations and 
relaxations~\cite{Lisowski:2005ey,Cavalieri:2007de,Rohwer:2011fy,Neppl:2015jh,Bertoni:2016ur}.
Resolving the spin polarization of photoelectrons with spin ARPES (sARPES) provides additional
information on the spin character of the sample~\cite{Hsieh:2009dp,Hsieh:2009if,Xu:2011fa,Jozwiak:2013kz}.
Exploring these new dimensions offer unprecedented opportunities to test our current understanding of matter.

Currently, the most common theoretical approaches to calculate ARPES are based on the one 
step-model where electron photoemission is treated as a unique coherent process 
that include all the scattering events~\cite{Pendry:1976io}. 
This is in contrast to the simpler three-step model where ionization is divided into three separate 
processes~\cite{Berglund:1964hq}.
These approaches, largely based on many-body perturbation theory formulated in terms of Green's function, 
proved to be successful in many relevant cases~\cite{Mahan:2000vt,Korringa:1947hb,Kohn:1954fh,Ebert:2011di}.
However the perturbative approach underlying these methods is not suitable to describe tARPES.\cite{Uimonen:2014cw} 
Including the time degree of freedom needed for tARPES requires a real time approach which, in the many-body 
context, is provided only by Keldysh Green's function theory. Efforts in this direction are still scarce and 
largely reduced to applications with model-Hamiltonians~\cite{Moritz:2010ko,Sentef:2013ks}, 
and current attempts to formulate the problem under the one-step model are still at the formal level~\cite{Braun:2015bs}.
 
In this paper we propose a completely different and computationally efficient approach based on a real-space real-time formulation of 
TDDFT where we obtain the ARPES spectra by directly analyzing 
the photoelectron current flux through a surface 
using the time-dependent surface flux method (t-SURFF).
The t-SURFF method is a well established technique that has been successfully employed to study ionization of atoms and small molecules under strong laser fields~\cite{Tao:2012ev,Scrinzi:2012jt,Majety:2015jr,Majety:2015fn,Zielinski:2015gg,Zielinski:2016kl}. This method has only recently been exported 
to TDDFT by some of the authors~\cite{Wopperer:2016ur}.
The application of t-SURFF to surfaces ionization dynamics is completely new and in this paper we 
present the extension of the method to semi-periodic systems. 
We name the extended method t-SURFFP. 
The resulting approach is fully ab-initio and capable to describe situations with any number 
of laser fields without making any assumption 
on the ionization process and at a modest computational cost.
For these reasons is naturally suited to simulate simulate tARPES.
Furthermore, in the spirit of the one-step model, scattering and surface effects are automatically 
included in the formalism.

The paper is organized as follows. 
In Sec.~\ref{sec:theory} we present our method in the context of TDDFT and in the more general 
context of ARPES. We then proceed to validate our technique in Sec.~\ref{sec:applications} where we
illustrate three representative cases: graphene, WSe$_2$, and hBN.
Finally, in Sec.~\ref{sec:conclusions} we discuss our findings and present the conclusions.

Atomic units ($\hbar=e=1$) are used throughout the paper unless otherwise specified.

\section{Theory} 
\label{sec:theory}

\subsection{General} 
\label{sub:general}
We here below briefly introduce the key concepts and quantities commonly used in the field
and that we will use thought the paper.

\begin{figure}
  \includegraphics[width=\linewidth]{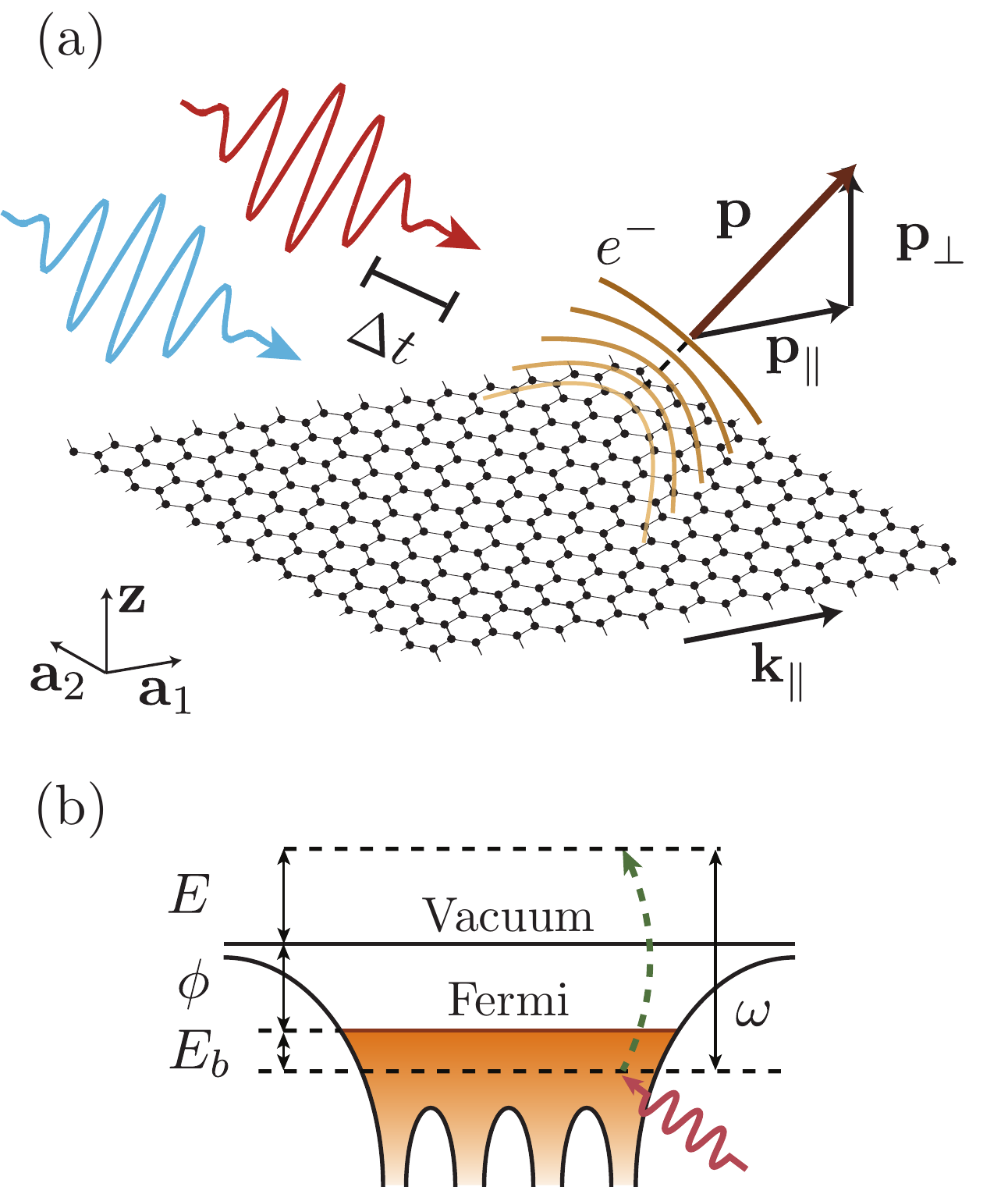}
  \caption{
    \label{fig:scheme_general}
    Schematic cartoon for the basic ingredients of tARPES. 
    In (a) we schematically illustrate the electron photoemission 
    process from a surface which periodically repeats itself along the lattice vectors $\mathbf{a}_1$ and
    $\mathbf{a}_2$.
    A useful picture to guide in the interpretation of ARPES is provided by the 
    photoelectric effect depicted in (b).
    }
\end{figure}

ARPES experiments are based on electron photoemission. 
When a field of appropriate energy $\omega$ irradiates a material surface a fraction of the electrons, 
originally bind to the crystal, is released into the vacuum as shown in Fig~\ref{fig:scheme_general} (a). 
These ionized electrons emerge with a kinetic energy distribution that depends both on the external field and
the material's electronic properties. 
Further away, at the detector, the kinetic energy distribution is measured as a function of energy $E$ and 
angle $\theta$, to form the angular resolved photoelectron spectrum $\mathcal{P}(E,\theta)$. 
This spectrum can be equivalently expressed as
a function of the escaping vector momentum $\mathbf{p}$, $\mathcal{P}(\mathbf{p})$. In some experiment a second stage 
is also capable to characterize the spin polarization along an arbitrary axis $\mathbf{s}$ and thus to measure 
$\mathcal{P}_{\mathbf{s}}(\mathbf{p})$.
Finally, time resolution can be achieved employing two pulses delayed by a time $\Delta t$ in a pump-probe setup.
The time resolved spectrum, $\mathcal{P}(\mathbf{p},\Delta t)$, is thus obtained by composing 
the spectra stroboscopically measured at different delays.

Einstein's photoelectric effect~\cite{Einstein:1905xx} underlies the interpretation of the physical information contained in the 
photoelectron spectrum represented schematically in Fig~\ref{fig:scheme_general} (b).
In fact, the kinetic energy distribution  of the electrons 
escaping the material follows the energy conservation relation $E=\omega -\phi-E_b$; where $\phi>0$ 
is the work function --
the minimum energy required to promote one electron into the vacuum -- and $E_b>0$ is the binding energy -- the 
band energy of the electrons relative to the Fermi level. To extract one electron from the material
requires a field with $\omega > \phi$. 

From energy conservation is thus apparent that the photoelectron spectrum contains information on the energy 
levels of the system. 
If the laser field is weak enough we can assume that momentum of the electron parallel to the surface $\mathbf{k}_\parallel$ -- the 
crystal momentum -- is conserved during the ionization process. This means that at the detector
$\mathbf{p}_\parallel=\mathbf{k}_\parallel$ and thus that $\mathcal{P}(\mathbf{p}_\parallel,E)$, obtained from 
ARPES $\mathcal{P}(\mathbf{p})$ with $E=\mathbf{p}^2/2$, directly maps to the electron dispersion in 
the crystal, i.e. the band structure.

This is the cornerstone of ARPES but is important to bear in mind that it is a strong idealization, 
useful to interpret the data, while in the reality of the experiment the picture can be more complex.
Several effects present in real materials may contribute to a departure from this picture. 
Among the most common we mention dynamical coupling to bosonic excitations such as phonons or 
plasmons~\cite{Grigorenko:2012jg,Lischner:2013fy,Caruso:2015dy} and surface effects~\cite{Hofer:1997cr, Anonymous:2000vt}.

In this paper we take an atomistic approach to simulate the photoemission process and include 
in the calculation the portion of space centered around the surface, both extending in the material 
and in the vacuum, that is needed to describe the process.
In the next section we describe the formalism, in the framework of TDDFT, to perform such a 
simulation and later discuss how ARPES can be obtained from the time-dependent density.

\subsection{Time-dependent spin-density functional theory for semi-periodic systems}
\label{sec:tdsdft}

In this work we describe systems with non-trivial spin configurations
derived from the presence of spin-orbit coupling (SOC) and therefore we use 
spin-density functional theory (SDFT)~\cite{vonBarth:1972eq}.

In SDFT the fundamental variable is the $2\times 2$ spin-density matrix
$\brho(\mathbf{r})=\rho_{\alpha\beta}(\mathbf{r})$ 
where the greek indices span the spin space $\alpha = +, -$.
This matrix is defined in terms of the spinless density -- or charge density -- $n(\mathbf{r})$ and 
the magnetization vector $\mathbf{m}(\mathbf{r})$ as follows
\begin{equation}\label{eq:rhomat}
\brho(\mathbf{r})= 
\frac{1}{2}n(\mathbf{r})\sigma_0+\frac{1}{2}\mathbf{m}(\mathbf{r})\cdot\boldsymbol{\sigma}  
\end{equation}
with $\boldsymbol{\sigma}=(\sigma_x,\sigma_y,\sigma_z)$ being the $2\times 2$
Pauli matrices and $\sigma_0$ the identity matrix.
From the definition it directly follows that the spinless density $n(\mathbf{r})$ can be 
obtained by tracing the spin-density matrix over the spin dimension $n(\mathbf{r})=\Tr(\brho(\mathbf{r}))$.

In this work we address the modeling on electrically driven systems for which the use of TDDFT is 
justified, for the case of time dependent magnetic fields one should in principle use TD-current 
DFT~\cite{Giuliani:826125}.
The central principle of TDDFT is that all observables of a time-dependent 
many-body system can be obtained from the knowledge of its time-dependent density alone
\cite{Runge:1984us,Marques:2011ud,cptddtf:2011}.
Likewise in static DFT the system of interacting particles is mapped into 
an auxiliary non-interacting system   
having the same time-dependent density, the Khon-Sham (KS) system~\cite{kohn1999}.
The KS system is represented by a Slater determinant composed of two-component spinors 
\begin{equation}
\bar{\varphi}_j(\mathbf{r})=\left[ \begin{array}{c}    \varphi_{j+}(\mathbf{r})\\ 
\varphi_{j-}(\mathbf{r})   \end{array}  \right]   
\end{equation}
whose time evolution is governed by the following time-dependent KS equations (TDKS) 
\begin{equation}\label{eq:tdks}
  \left\{
\begin{aligned}
   i&\frac{\partial  }{\partial t}\bar{\varphi}_j (\mathbf{r},t) = \hat{H}_{\rm KS}[\brho](\mathbf{r}) \bar{\varphi}_j(\mathbf{r},t) \\
  &\hat{H}_{\rm KS}[\brho](\mathbf{r}) = \\ 
   & -\frac{1}{2}\left(\boldsymbol{\nabla}-\frac{\mathbf{A}(t)}{c}\right)^2 \sigma_0
  + V_{\rm ion}(\mathbf{r})+ V_{\rm KS}[\brho](\mathbf{r})
\end{aligned}  
\right.
\end{equation}
with $\hat{H}_{\rm KS}[\brho](\mathbf{r})$ being the KS Hamiltonian composed of the external laser field expressed as a time-dependent vector potential in the velocity gauge $\mathbf{A}(t)$ (with the electric field being $\mathbf{E}(t)=\partial \mathbf{A}/\partial t$), the external potential 
generated by the ions in the lattice $V_{\rm ion}(\mathbf{r})$  and 
the KS potential $V_{\rm KS}[\brho](\mathbf{r})$. 
The KS potential 
is the sum of the classical electrostatic potential $V_{\rm H}[n](\mathbf{r})$, that only depends on 
the spinless density, and the exchange and correlation potential $V_{\rm xc}[\brho](\mathbf{r})$
responsible for the many-body interaction
\begin{equation}\label{eq:vks}
V_{\rm KS}[\brho](\mathbf{r}) = V_{\rm H}[n](\mathbf{r}) + V_{\rm xc}[\brho](\mathbf{r}) \,.
\end{equation}
Non-trivial spin configurations are induced by spin-orbit coupling (SOC) whenever heavy ions are present in the crystal.
In practice this introduces a term proportional to $\mathbf{L}\cdot \mathbf{S}$ in 
the ions' potential $V_{\rm ion}(\mathbf{r})$ that breaks spin rotational symmetry
and allows for non-collinear spin configurations.

The KS Hamiltonian in~\eqref{eq:tdks} has a functional dependence on the spin-density matrix $\brho$ which 
can be reconstructed from the spinors using \eqref{eq:rhomat} and 
the charge density and magnetization vector defined by 
\begin{eqnarray}\label{eq:nmr}
n(\mathbf{r})&=&\sum_{j=1}\theta (\mu 
-\epsilon_j)\bar{\varphi}_j(\mathbf{r})^\dagger \bar{\varphi}_j(\mathbf{r}) \label{eq:nr} \\
\mathbf{m}(\mathbf{r})&=& 
\sum_{j=1} \theta (\mu -\epsilon_j) \bar{\varphi}_j(\mathbf{r})^\dagger \boldsymbol{\sigma} 
\bar{\varphi}_j(\mathbf{r}) \label{eq:mr}
\end{eqnarray}
where $\epsilon_j$ is the $j$-th eigenvalue of $\hat{H}_{\rm KS}[\brho](\mathbf{r})$ and $\mu$ is the Fermi level  
obtained with the constraint that the charge density integrates to the total number of electrons: 
$N=\int {\rm d}\mathbf{r} n(\mathbf{r})$.

In general $\hat{H}_{\rm KS}[\brho](\mathbf{r})$ is not diagonal in spin. The only 
diagonal term is the gauge-invariant kinetic operator -- the first term of KS Hamiltonian in~\eqref{eq:tdks}. However, in crystals composed of light atoms, for which SOC is negligible, and in  absence of a magnetic field the spin-density is collinear. 
This means that $\mathbf{m}(\mathbf{r})$ is constant in space and that is 
possible to choose a reference frame where $\hat{H}_{\rm KS}[\brho](\mathbf{r})$ is always diagonal. 
In this case we can simplify the formalism decoupling spin up and spin down 
to obtain two separate set of equations -- the spin-polarized TDKS equations.

\begin{figure}
  \includegraphics[width=\linewidth]{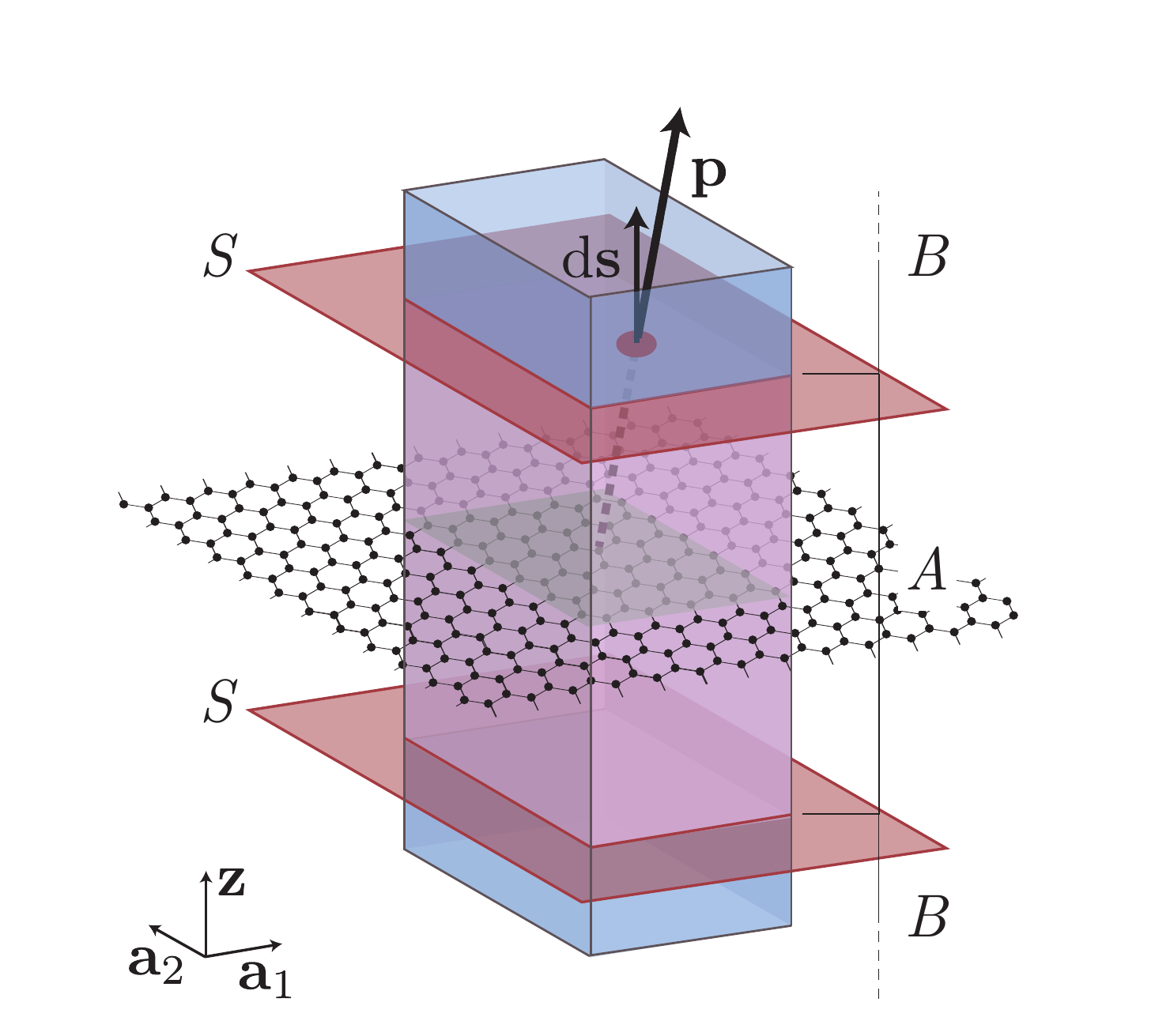}
  \caption{
    \label{fig:scheme}
    Scheme illustrating the geometrical model needed to simulate the ionization 
    process in a semi-periodic system. 
    }
\end{figure}

In order to describe photoemission processes from first principles we have to describe the interface 
between the material and the vacuum. 
To this end we model a surface as a semi-periodic structure repeating itself along two directions identified by the 
lattice vectors $\mathbf{a}_1$ and $\mathbf{a}_2$ like in Fig.~\ref{fig:scheme}. 
In the figure we indicate with $\mathbf{z}$ the non-periodic dimension. 

Owing to the periodicity of the system we can describe the infinite surface with wavefunctions
confined to the volume $\Omega=\mathbf{a}_1 \times \mathbf{a}_2 \times \mathbb{R}$ where 
 $\mathbf{a}_1$ and $\mathbf{a}_2$ span the planar primitive cell.
On the surface, we express the wavefunctions as Bloch spinors $\bar{\varphi}_{j\mathbf{k}}(\mathbf{r})=e^{i 
\mathbf{k}\cdot\mathbf{r}} \bar{u}_{j\mathbf{k}}(\mathbf{r})$ where $\bar{u}_{j\mathbf{k}}(\mathbf{r})$ 
is a spinor with the periodicity of the lattice and $\mathbf{k}$ covers the first Brillouin 
zone (BZ) of the reciprocal space, which is bidimensional.
Using  Bloch spinors corresponds to exchanging
\begin{equation}
\left(\boldsymbol{\nabla}-\frac{\mathbf{A}(t)}{c}\right)^2 \rightarrow  
\left(\boldsymbol{\nabla}+i\mathbf{k} -\frac{\mathbf{A}(t)}{c} \right)^2
\end{equation}
in the KS Hamiltonian, $\hat{H}_{\rm KS}[\brho](\mathbf{r})$, defined in~\eqref{eq:tdks}.
The KS equations describing the infinite surface is then cast into a set of equations for each 
value of $\mathbf{k}$ coupled trough $\boldsymbol{\rho}(\mathbf{r})$ which can be obtained from 
\eqref{eq:rhomat} integrating the Bloch spinors over the first BZ in Eq.~\eqref{eq:nr} and 
\eqref{eq:mr}.

\subsection{The t-SURFFP method}

To derive a suitable formalism for photoemission with TDDFT in semi-periodic systems
we employ the t-SURFF method~\cite{Tao:2012ev,Scrinzi:2012jt}. 
This method was recently extended to TDDFT in finite systems ~\cite{Wopperer:2016ur} 
and we hereby present a further extension to the semi-periodic case that we name t-SURFFP.

To this end we partition the volume $\Omega$ along the non-periodic dimension in two regions, $A$ and $B$, 
as illustrated in Fig.~\ref{fig:scheme}. 
We assume that in region $A$ electrons are fully interacting and described by the KS Hamiltonian 
$\hat{H}_{\rm KS}(t)$ while in $B$ they are non-interacting and free. 
In other words we ask that the time-dependent Hamiltonian governing the evolution of the system, 
$\hat{H}(t)$, asymptotically reduces to an exactly solvable one, 
$\hat{H}_{\rm V}(t)$, for all times
\begin{equation}\label{eq:h_partition}
  \hat{H}(t)=\left\{
    \begin{aligned}
    \hat{H}_{\rm KS}(t) \quad &{\rm for}\, \mathbf{r}\in A \\
    \hat{H}_{\rm V}(t)  \quad &{\rm for}\, \mathbf{r}\in B         
    \end{aligned}
  \right. .
\end{equation}
In our scheme $\hat{H}_{\rm V}(t)$ is the Volkov Hamiltonian governing the dynamics of 
$N$ non-interacting free electrons in $\Omega$ driven by a time-dependent external 
field $\mathbf{A}(t)$; 
in the velocity gauge this is expressed by
\begin{equation}
  \hat{H}_{\rm V}(t) =\sum_{j=1}^N \frac{1}{2}\left[-i\nabla_j -\frac{\mathbf{A}(t)}{c}\right]^2\sigma_0\,,
\end{equation}
and is diagonal in spin-space as indicated by the presence of $\sigma_0$.
Provided the vector field $\mathbf{A}(t)$ is constant in space we can solve exactly the 
time dependent Scr\"odinger equation associated with $\hat{H}_{\rm V}(t)$.  
For each single electron the solution can be written in the form
of a plane wave spinor with momentum $\mathbf{p}$
\begin{equation}\label{eq:volkov_wf}
  \bar{\chi}_{\mathbf{p}}(\mathbf{r},t)=\sqrt{\frac{2\pi} {a_1 a_2}} e^{i \mathbf{p}\cdot \mathbf{r}} e^{-i\phi(\mathbf{p},t)}
\end{equation}
multiplied by a time-dependent phase factor
\begin{equation}
  \phi(\mathbf{p},t)=\frac{1}{2}\int_0^t {\rm d}\tau  \left[\mathbf{p} -\frac{\mathbf{A}(\tau)}{c}\right]^2  \,.
\end{equation}
Each spinor is normalized on $\Omega$ which is finite 
along $\mathbf{a}_1$ and $\mathbf{a}_2$ but infinite along $\mathbf{z}$, and the
normalization factor precisely accounts for this geometry. 
Further, owing to periodic boundary conditions along $\mathbf{a}_1$ and $\mathbf{a}_2$, we can  
decompose $\mathbf{p}$ into a sum of a k-point $\mathbf{k}$, 
bound to the plane of the surface, and a general reciprocal lattice vector $\mathbf{G}$:
$\mathbf{p}=\mathbf{k}+\mathbf{G}$. 
Owing to the periodicity of the system $\mathbf{G}_\parallel$ assumes discrete values while 
$\mathbf{G}_\perp$ is continuos.
Since the wavefunctions in~\eqref{eq:volkov_wf} are not pure Volkov waves but retain information 
about the periodic dimensions we denote them as Bloch-Volkov waves.

When the system ionizes we can make a further assumption on the spatial distribution of the 
wavefunction.
In the long time limit, after the external field has been switched of $\mathbf{A}(t>T)=0$, we  
assume that each KS spinor is factorizable into a bound and a scattering component localized 
in $A$ and $B$ respectively,
\begin{equation}\label{eq:wf_partition}
\bar{\varphi}_{j\mathbf{k}}(\mathbf{r},t)=\bar{\varphi}_{j\mathbf{k},A}(\mathbf{r},t)+\bar{\varphi}_{j\mathbf{k},B}(\mathbf{r},t) \,.
\end{equation}

Under this assumption the number of electron escaped per unit cell from $A$ at time $T$ can be expressed as
\begin{align}\label{eq:n_esc}
  N_{\rm esc}(T)&= \int\limits_\Omega {\rm d}r \,n_B(\mathbf{r},T) \\
                &=\int\limits_\Omega {\rm d}r \int\limits_{BZ} {\rm d}\mathbf{k} 
                \sum_{j=1} \theta_j  |\bar{\varphi}_{j\mathbf{k},B}(\mathbf{r},T)|^2\nonumber
\end{align}
where $\theta_j$ is a shorthand for $\theta (\mu-\epsilon_j)$ 
as in~\eqref{eq:nmr}, and $n_B(\mathbf{r},T)$ is the charge density in $B$.

Since the Coulomb-Volkov waves form a complete set we can expand each KS spinor as 
\begin{equation}\label{eq:vw_expansion}
  \bar{\varphi}_{j\mathbf{k},B}(\mathbf{r},t) = \int {\rm d}p\,  \bar{b}_j(\mathbf{p})\chi_{\mathbf{p}}(\mathbf{r},t)\,,
\end{equation}
where we defined the coefficients 
\begin{equation}
\bar{b}_j(\mathbf{p})\equiv\left[ \begin{array}{c} b_{j+1/2}(\mathbf{p})\\ b_{j-1/2}(\mathbf{p})   \end{array}\right] 
\end{equation}
as column vectors in spin space.
Inserting \eqref{eq:vw_expansion} into \eqref{eq:n_esc} we obtain that the number of escaped electrons 
can be expressed in terms of the expansion coefficients $b_{j\alpha}(\mathbf{p})$ 
by tracing over the spin components as follows
\begin{equation}\label{eq:nesc_bip}
  N_{\rm esc}(T) = \sum_{j=1} \theta_j  \sum_{\alpha=-}^{+} \int\limits_{BZ} {\rm d}\mathbf{k}  \int {\rm d}\mathbf{p} \,|b_{j\alpha}(\mathbf{p})|^2\, .
\end{equation}
The spinless momentum-resolved photoelectron probability $\mathcal{P}(\mathbf{p})$ is thus naturally 
obtained from the former expansion as the derivative with respect to $\mathbf{p}$ of $N_{\rm esc}(T)$
\begin{equation}\label{eq:diff_cross}
  \mathcal{P}(\mathbf{p})=\frac{\partial N_{\rm esc}(T)}{\partial \mathbf{p}}=\sum_{j=1} \theta_j  \sum_{\alpha=-}^{+}\int\limits_{BZ} {\rm d}\mathbf{k}\, |b_{j\alpha}(\mathbf{p})|^2 .
\end{equation}
In order to calculate $\mathcal{P}(\mathbf{p})$ we thus need an explicit form for the
expansion coefficients compatible with a TDDFT formulation. 
This is provided by the flux of the photoelectron current through a closed surface. 

Using the continuity equation we can express the number of escaped electrons as the flux of the 
current density $\mathbf{J}(\mathbf{r},t)$ trough a surface $S$ enclosing the system. 
By choosing $S$ as in Fig.~\ref{fig:scheme} parallel to the 
system's plane we have that 
\begin{equation}\label{eq:nesc_flux}
  N_{\rm esc}(T) = -\int\limits_0\limits^T {\rm d}\tau \oint\limits_S {\rm d}\mathbf{s} \cdot \mathbf{J}(\mathbf{r},\tau)\, .
\end{equation}
We are thus left with the task of connecting $\mathbf{J}(\mathbf{r},t)$ with the KS spinors. 
This is achieved observing that $\mathbf{J}(\mathbf{r},t)$ can be expressed 
as the expectation value of the single particle current density operator 
\begin{equation}
  \hat{\mathbf{j}}(t)=\frac{1}{2}\left[ \left( -i\nabla -\frac{\mathbf{A}(t)}{c}\right) 
  + {\rm c.c.}\right]\sigma_0
\end{equation}
over KS orbitals as follows
\begin{equation}
  \mathbf{J}(\mathbf{r},t) =  \sum_{j=1} \theta_j  \int\limits_{BZ} {\rm d}\mathbf{k} 
  \langle \bar{\varphi}_{j\mathbf{k}}(t)|\hat{\mathbf{j}}(t)| \bar{\varphi}_{j\mathbf{k}}(t) \rangle . 
\end{equation}
We can then use \eqref{eq:vw_expansion} to expand the bra in the former equation to obtain
\begin{equation}\label{eq:current_bip}
  \mathbf{J}(\mathbf{r},t) =  \sum_{j=1} \theta_j  \int\limits_{BZ} {\rm d}\mathbf{k} \int {\rm d}\mathbf{p}\, \bar{b}_j^*(\mathbf{p})
  \langle \chi_{\mathbf{p}}(t)|\hat{\mathbf{j}}(t)| \bar{\varphi}_{j\mathbf{k}}(t) \rangle
\end{equation}
and the complex conjugated counterpart by expanding the ket
\begin{equation}\label{eq:current_bip_cjg}
  \mathbf{J}(\mathbf{r},t) =  \sum_{j=1} \theta_j  \int\limits_{BZ} {\rm d}\mathbf{k} \int {\rm d}\mathbf{p}\, \bar{b}_j(\mathbf{p})
  \langle  \bar{\varphi}_{j\mathbf{k}}(t)|\hat{\mathbf{j}}(t)|\chi_{\mathbf{p}}(t) \rangle\,.
\end{equation}
Finally, by inserting \eqref{eq:current_bip} and \eqref{eq:current_bip_cjg} into 
\eqref{eq:nesc_flux} and directly comparing the resulting equations with \eqref{eq:nesc_bip} we 
arrive at an explicit equation for the Bloch-Volkov expansion coefficients in the form of flux 
integral
\begin{equation}
  \bar{b}_j(\mathbf{p}) = -\int\limits_0\limits^T {\rm d}\tau \oint\limits_S {\rm d}\mathbf{s} 
  \cdot\langle \chi_{\mathbf{p}}(\tau)|\hat{\mathbf{j}}(\tau)| \bar{\varphi}_{j\mathbf{k}}(\tau) \rangle \,.
\end{equation}
Further, by exposing the k-point dependence of each Bloch-Volkov spinor, we can recast the 
previous equation into a form containing only the periodic component of each KS 
orbital as 
\begin{equation}\label{eq:bjp_flux}
  \bar{b}_j(\mathbf{p}) = -\int\limits_0\limits^T {\rm d}\tau \oint\limits_S {\rm d}\mathbf{s} 
  \cdot\langle \mathbf{G}|\hat{\mathbf{j}}(\tau)| \bar{u}_{j\mathbf{k}}(\tau) \rangle e^{i\phi(\mathbf{p},\tau)}\, ,
\end{equation}
where $| \mathbf{G} \rangle $ are planewaves of momentum $\mathbf{G}$ normalized in $\Omega$ as in Eq.~\eqref{eq:volkov_wf}. 
This formulation is particularly convenient for numerical implementations since it involves
only the periodic spinors $\bar{u}_{j\mathbf{k}}(\tau)$ and thus it fully exploits the Bloch 
factorization of the KS equations discussed in Sec.~\ref{sec:tdsdft}.

Equation \eqref{eq:bjp_flux} together with \eqref{eq:diff_cross} provide a straightforward 
way to calculate the spinless momentum-resolved photoelectron probability $\mathcal{P}(\mathbf{p})$.
We recall that this result has been derived under the assumption that, 
(i) the Hamiltonian of the system can be well approximated with \eqref{eq:h_partition}, and (ii) 
that scattering and bound electrons are spatially well separated at all times 
\eqref{eq:wf_partition}.
In addition, by choosing TDDFT as working framework, we assumed that the longitudinal 
part of the photoelectron current is the one that contribute the most in the photoelectron spectrum. 
These conditions ultimately define the range of applicability of the method. 
They are clearly satisfied at an infinite distance from the system and close to the detectors
where electrons can be safely considered free forward-moving particles, 
and poor in the vicinity of the surface. 
Positioning the sampling surface $S$ is the only parameter controlling 
the accuracy of this approximation and, in practical calculations, it has to be
varied until a convergence of the spectrum is achieved.

Once the expansion coefficients are calculated we can obtain the spin-density photoelectron 
probability $\mathcal{P}_{\alpha\beta}(\mathbf{p})=\mathbf{P}(\mathbf{p})$ 
with a procedure similar to the one employed to construct the spin-density matrix of 
Eq.~\eqref{eq:rhomat}, i.e. by simply substituting the spinors in the definitions of \eqref{eq:nr} and \eqref{eq:mr} with $\bar{b}_{j}(\mathbf{p})$ and then use equation Eq.~\eqref{eq:rhomat}.
The spin-resolved photoelectron probability polarized along the direction $\mathbf{s}$ is 
obtained tracing over the spin dimension as follows
\begin{equation}
\mathcal{P}_{\mathbf{s}}(\mathbf{p})= 
\Tr[\mathbf{P}(\mathbf{p}) \boldsymbol{\sigma}\cdot\mathbf{s}]\,.  
\end{equation}
The spinless photoelectron probability can also be obtained tracing out the spin degrees of freedom
consistently with~\eqref{eq:diff_cross}: $\mathcal{P}(\mathbf{p})= \Tr[\mathbf{P}(\mathbf{p})]$.
The photoelectron spin polarization is then defined as the ratio between spin-resolved and spinless 
probabilities 
\begin{equation}\label{eq:pis}
\Pi_\mathbf{s}(\mathbf{p})=\frac{\mathcal{P}_{\mathbf{s}}(\mathbf{p})}{\mathcal{P}(\mathbf{p})}\,.  
\end{equation}

From the knowledge of $\mathcal{P}(\mathbf{p})$, ARPES is directly obtained by 
looking at the electron photoemission probability as a function of the total kinetic energy $E$ and
the parallel component of the escaping momentum $\mathbf{p}_\parallel$ as discussed in 
Sec.~\ref{sub:general}: 
$\mathcal{P}(\mathbf{k}_\parallel=\mathbf{p}_\parallel, E=\frac{\mathbf{p}^2}{2})$.

Note that the external filed $\mathbf{A}(t)$ we use to derive the t-SURFFP equation~\eqref{eq:bjp_flux} 
and to propagate the KS equations~\eqref{eq:tdks} is free to assume any arbitrary temporal shape. 
This means that it can describe any linear combination of laser fields. 
Our approach thus provides a straightforward environment to simulate tARPES where the time variable is 
extracted trough the variation of a time-delay between two laser pulses in a pump-probe configuration.

\section{Applications} 
\label{sec:applications}

\subsection{Computational details}

We implemented the t-SURFFP method in the Octopus code~\cite{Marques:2003wr,PSSB:PSSB200642067,Strubbe:2015iz}.
In Octopus the TDKS equations are solved in a real-space grid and propagated in real-time. 
The real-space approach offers a great versatility when it comes to 
the description of semi-periodic systems as it naturally allows to impose mixed boundary 
conditions. 
The real-time propagation offers additional flexibility as it consents to describe external fields 
with arbitrary temporal shape.
In our implementation we used the exact cutoff method described in Ref.~\cite{Rozzi:2006js} along the non-periodic dimension and employed non-orthogonal grids~\cite{Natan:2008ka}, optimized according to the lattice symmetries, on the periodic ones.
Our implementation is fully parallel in grid points, k-points, bands and spin dimension. 
For the largest systems presented in this paper (bilayer WSe$_2$), we found that the distribution 
over only k-points and states is enough to saturate a medium-size cluster ($\approx$ 1024 cores).

In the code we use a pseudopotential formalism where only valence electrons are treated explicitly.
Core electrons together with the ionic potential are replaced by an effective pseudopotential
such that the ionic potential is composed of a local potential, a non-local one plus a SOC term
\begin{equation}\label{eq:vion}
  V_{\rm ion}(\mathbf{r}) = V_{\rm local}(\mathbf{r}) + V_{\rm nlocal}(\mathbf{r}) + V_{\rm SO}\mathbf{L}\cdot 
  \mathbf{S}\,.
\end{equation}
In all the calculations we used HGH pseudopotentials~\cite{Hartwigsen:1998dk} 
accounting for relativistic effects only for the compounds containing W.

Further, we used the local density approximation (LDA)~\cite{Perdew:1981dv} to the exchange and correlation functional.
For non-collinear spin configurations we treat the functional at the level of local spin-density 
approximation (LSDA) by rotating to the local reference frame where the spin density matrix is 
diagonal~\cite{Kubler:1988}.

In order to to prevent spurious reflection on the non-periodic edges of the cell we 
employed complex absorbing-potential boundary conditions~\cite{Larsen:2015kc}. 
We tune the absorber parameters (width and imaginary amplitude) such that the boundary is 
effective in the energy window of the ejected photoelectrons~\cite{DeGiovannini:2015jt}.
Owing to the presence of absorbing boundaries the total charge in the cell is not conserved over 
time. 
To avoid artifacts from charge imbalance we chose the laser intensity such that the total charge 
loss is negligible.
In our calculations we found that a 10$^{-4}$\% of charge loss is sufficient to provide 
stable results.

In the following we present applications for three different materials having an hexagonal 
lattice.
These are all stable layered structures that are currently in the focus of extensive research and for
which a considerable amount of high quality experimental data is being produced. 
We stress however that the computational tool we developed is by no means restricted to hexagonal crystal 
lattices only.

In all the simulations we used laser pulses that are zero everywhere except for $t\in[0,T]$ with
\begin{equation}
  \mathbf{A}(t) = \boldsymbol{\epsilon} A_0\sin\left( \frac{\pi t}{T}\right)^2\cos(\omega t)\,;
\end{equation} 
here $\boldsymbol{\epsilon}$ is the polarization axis, $A_0$ is the peak amplitude and $\omega$ the 
carrier frequency.
This choice is motivated by the resemblance to the typical experimental Gaussian shape while retaining 
the property of being exactly zero outside a given time window -- this is important to minimize the 
propagation time in the simulations. 
All the simulations have been carried out on a box of 120~au along the non-periodic dimension 
and centered around each system. Along the periodic dimension the box is taken according to 
the primitive cell of the system. Complex absorbing potentials of 30~au width have been placed at 
the opposite sides of the simulation box to prevent reflections.  
The t-SURFFP analyzing surface was placed at 30~a.u. from the edges right before the onset of the 
absorbers. 
Finally the first BZ was sampled by a $12\times12$ grid of $\mathbf{k}$-points in reciprocal space
for all the systems.

\subsection{Graphene} 
\label{sub:graphene}

In this section we illustrate the application of t-SURFFP to simulate 
ARPES from graphene monolayer. 
To this end we ionize the system with a 50~fs laser pulse with with $\omega=95$~eV,
$\boldsymbol{\epsilon}=\mathbf{z}$, and peak intensity $I=10^9$~W/cm$^2$. 
In an experiment this geometry corresponds to the case where the laser is grazing 
with respect to the surface. 
In our calculations we used a grid spacing of 0.36~a.u. and a lattice constant of $a=4.65$~a.u..

\begin{figure}
  \includegraphics[width=\linewidth]{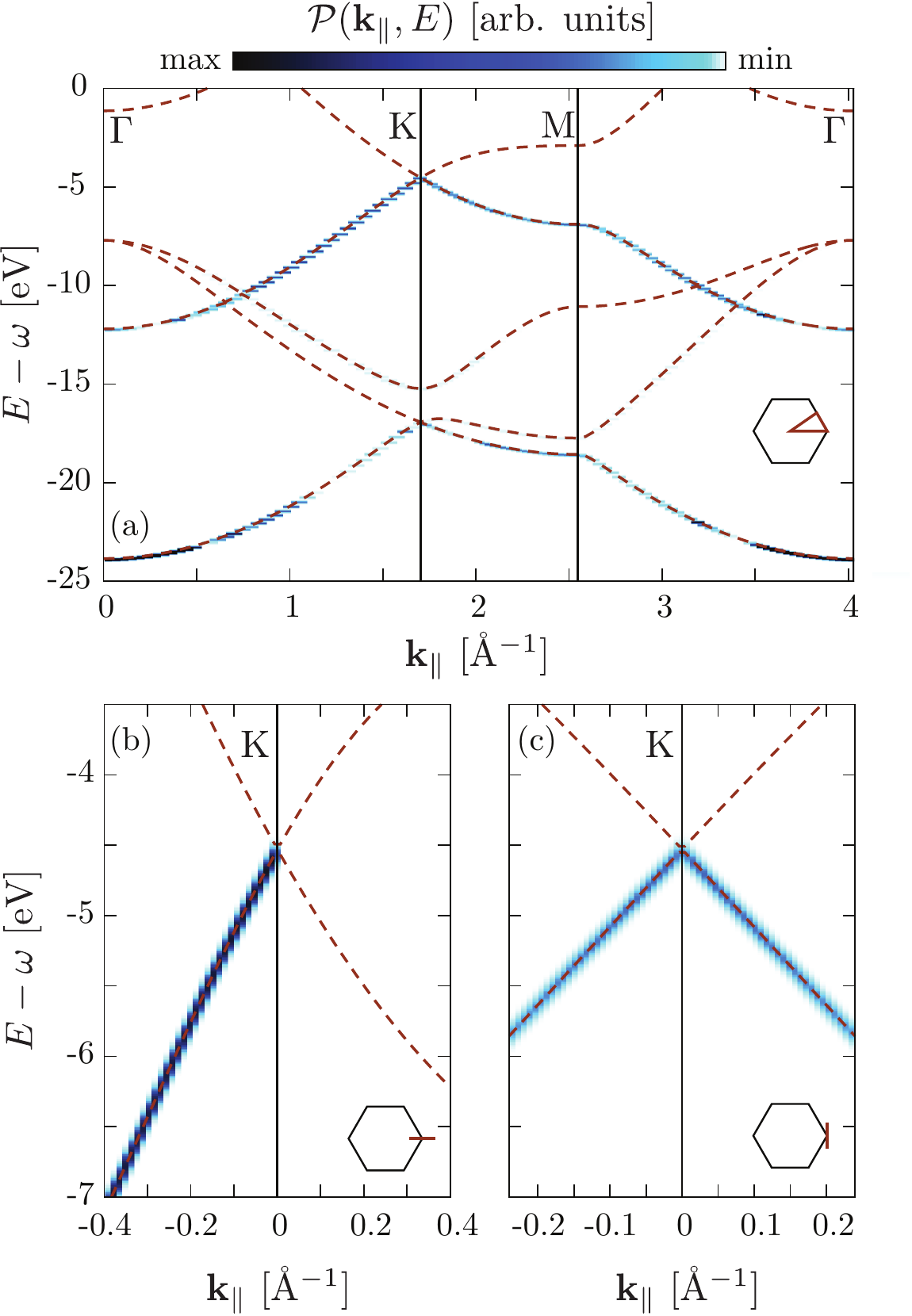}
  \caption{\label{fig:graphene_arpes}
  Graphene ARPES as obtained using a 50 fs pulse with $\omega=95$~eV,
  $\boldsymbol{\epsilon}=\mathbf{z}$, and peak intensity $I=10^9$~W/cm$^2$.
  (a) ARPES cut along the $\Gamma$-K-M-$\Gamma$ path in the BZ, 
  (b) cut along a direction parallel to $\Gamma$-K centered in K and (c)
  cut along a direction parallel to $\Gamma$-M also centered in K.
  In all panel the inset schematically represents the reciprocal space path on 
  which the spectra are plotted, and the DFT band structure is overlaid in red. 
    }
\end{figure}

The results of the simulation are illustrated in Fig.~\ref{fig:graphene_arpes}~(a) 
on the $\Gamma$-K-M-$\Gamma$ path in the BZ (see inset). 
ARPES presents intensity peaks that are positioned in excellent agreement with the DFT band structure (overlaid with red lines) and thus is in agrement with the interpretation of the 
ionization process in terms of the photoelectric effect.

Not all the DFT bands are visible in the spectrum. 
This behavior can be explained in first order time-dependent perturbation theory with Fermi's golden 
rule wich describes the probability to excite an initial state in the material $|i\rangle$ to final 
state in the continuum $|f\rangle$ in terms of the dipole matrix element 
$|\langle f|\hat{\mathbf{p}}\cdot \boldsymbol{\epsilon}| i\rangle|^2$.
Final states that are connected with negligible matrix elements appear dark in the spectrum. 
The matrix element intensity effects observed in the calculation are in excellent agreement with the 
literature~\cite{Bostwick:2006hn}.

Around the Dirac point, on the valence band close to K, graphene ARPES presents a peculiar 
intensity pattern. Only one branch of the Dirac cone is visible crossing K from a path along the $\Gamma$-K direction while both are visible from a direction parallel to $\Gamma$-M as shown in 
Fig.~\ref{fig:graphene_arpes}~(b) and (c). 
This is a characteristic feature of ARPES on graphene which has been observed in many experiments~\cite{Bostwick:2006hn}
and that is due to the chiral character of Dirac states at the K point.~\cite{MuchaKruczynski:2008jd}

In this work we do not include any dissipation channel. 
The finite line-width observed in ARPES is thus a direct consequence of the finite time window of the probe pulse. 
In principle, however, it could be possible to include dissipation, for instance, by coupling the electronics degrees of freedom with lattice vibrations using Ehrenfest theorem.\cite{Andrade:2009ga} 

\subsection{WSe$_2$} 

In this section we turn to a system with a non-trivial spin configuration 
and study photoemission from the transition metal dichalcogenide WSe$_2$.

\begin{figure}
  \includegraphics[width=\linewidth]{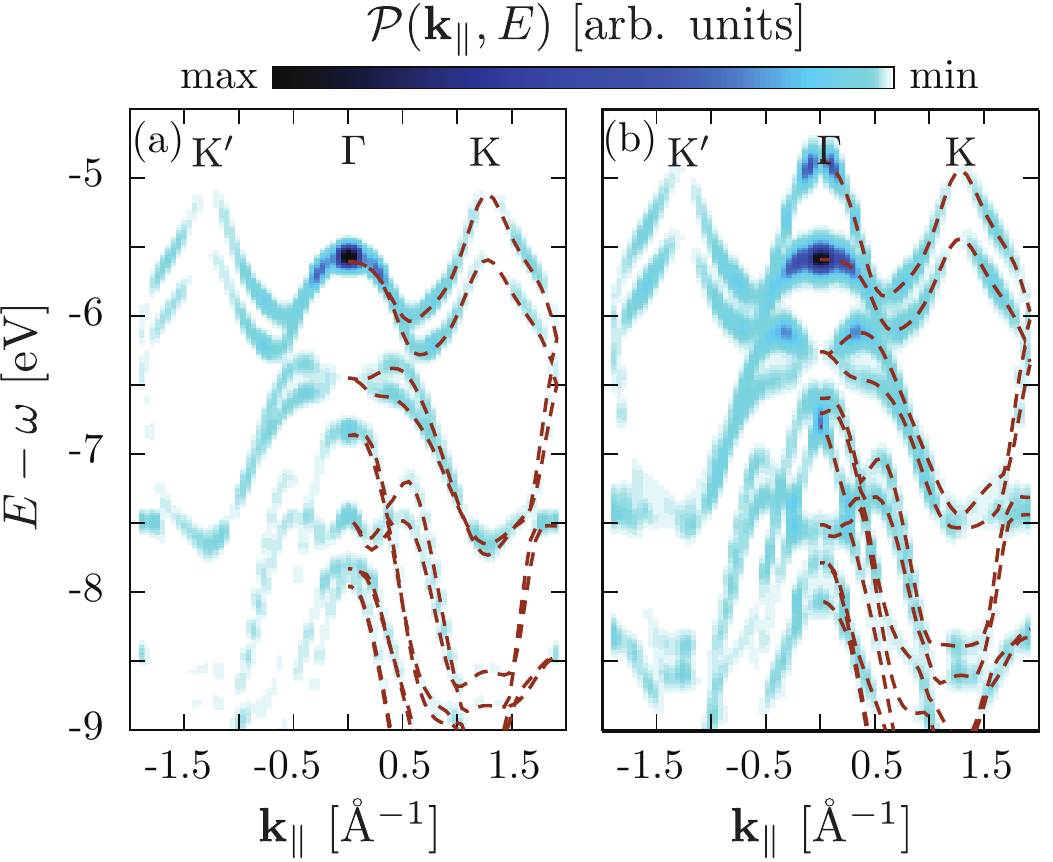}
  \caption{\label{fig:wse2_arpes}
    ARPES on WSe$_2$ monolayer (a) and bilayer (b). 
    To produce the spectra we used a 48 fs laser pulse with 
    frequency $\omega=127$~eV, polarization $\boldsymbol{\epsilon}=\mathbf{z}$, 
    and peak intensity $I=10^9$~W/cm$^2$.
    The ground state band structure is depicted in red. 
    }
\end{figure}

We probe the system with a 48~fs laser pulse polarized along 
$\mathbf{z}$ with carrier frequency $\omega=127$~eV and peak intensity $I=10^9$~W/cm$^2$. 
We performed the simulations with a grid spacing of 0.4~a.u., 
employed a lattice constant of $a=6.2$~a.u., and included semicore electrons in the pseudopotentail
for W.  
The results are presented in Fig.~\ref{fig:wse2_arpes} (a)
and (b) for monolayer and bilayer respectively. 

As observed in the previous section the agreement between the ARPES spectrum and the 
equilibrium band structure is excellent. 
In addition, the results for the bilayer in Fig.~\ref{fig:wse2_arpes}~(b) is in good agreement with ARPES 
experiments recently reported on bulk WSe$_2$~\cite{Riley:2014bw,Riley:2015fm}.
The reason for such agreement is related to the surface sensibility of ARPES experiments.\cite{Bertoni:2016ur} 
In fact, scattering prevents photoelectrons to be ejected from the lower lying layers of the material and effectively only the topmost layers at the surface contribute to the spectrum. 
By comparing this spectrum to the experiments reported in Refs.~\cite{Riley:2014bw,Riley:2015fm} it becomes clear that the system probed by the experiment is 
composed of more than two layers as indicated, for instance, by the presence of an ARPES signal filling the 
space between the two topmost valence bands at $\Gamma$.

\begin{figure}
  \includegraphics[width=\linewidth]{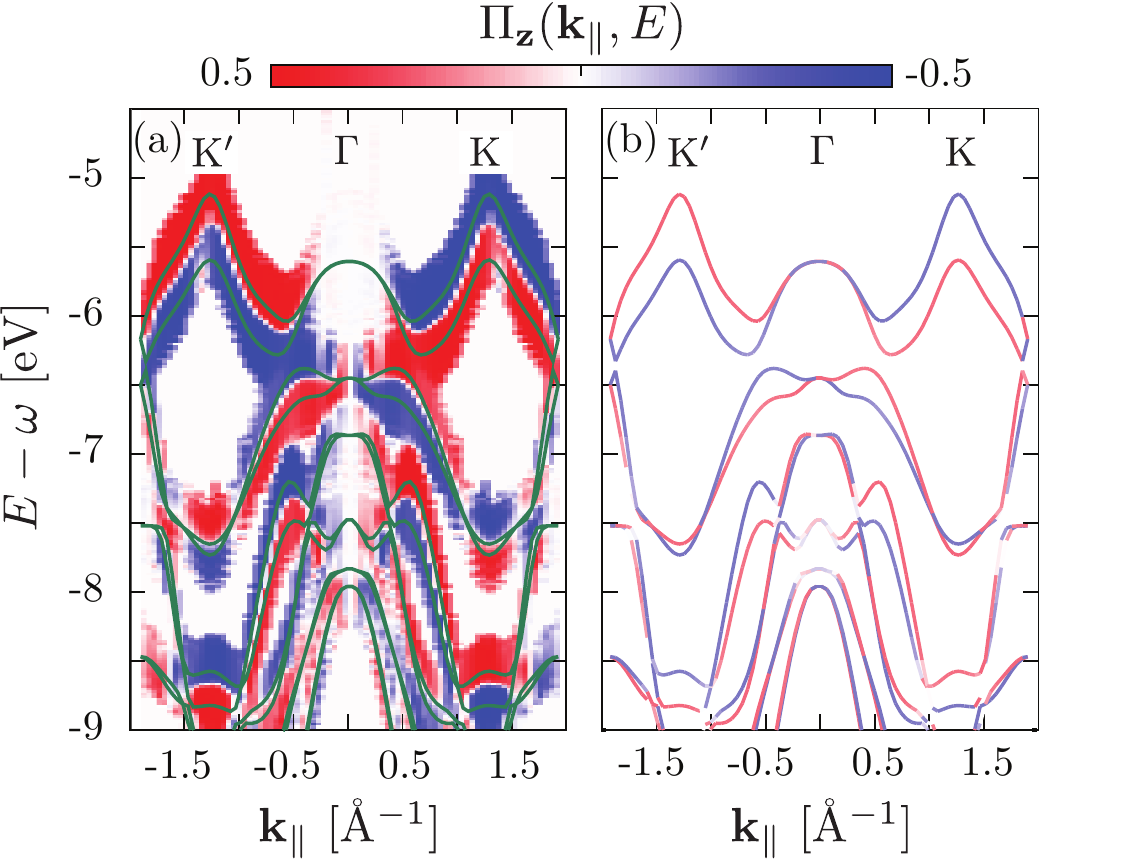}
  \caption{\label{fig:wse2sarpes}
    Spin-resolved ARPES for monolayer WSe$_2$. In panel (a) we show 
    $\Pi_\mathbf{z}(\mathbf{k}_\parallel,E)$, the photoelectron spin polarization 
    spectrum along $z$ overlaid with the band structure in green. 
    In (b) is depicted the spin polarization of the ground state band structure.
    Laser parameters are the same as in Fig.~\ref{fig:wse2_arpes}.
    }
\end{figure}

In monolayer WSe$_2$ inversion symmetry is broken. 
For this reason the high-symmetry points K and K$^\prime$ in reciprocal space are no longer equivalent. 
This fact combined with a strong SOC provides a large splitting and polarization of the bands which 
is opposite for K and K$^\prime$.
Both splitting and spin polarization can be measured with sARPES as illustrated 
in Fig.~\ref{fig:wse2sarpes}~(a).
According to sARPES the topmost valence bands at K and K$^\prime$ are fully spin polarized and 
with opposite spins. At $\Gamma$, where the bands are degenerate the spin polarization is zero. 
This behavior is consistent with the spin polarization of the DFT bands shown in 
Fig.~\ref{fig:wse2sarpes}~(a) where becomes apparent that the zero spin-polarization 
at $\Gamma$ emerges from degenerate bands with opposite polarization.

\subsection{Hexagonal Boron-nitride} 
\label{ssub:boron_itride}
In this section we explore the dimension offered by tARPES and illustrate 
the case of monolayer hBN in a pump and probe setup.

\begin{figure}
  \includegraphics[width=0.9\linewidth]{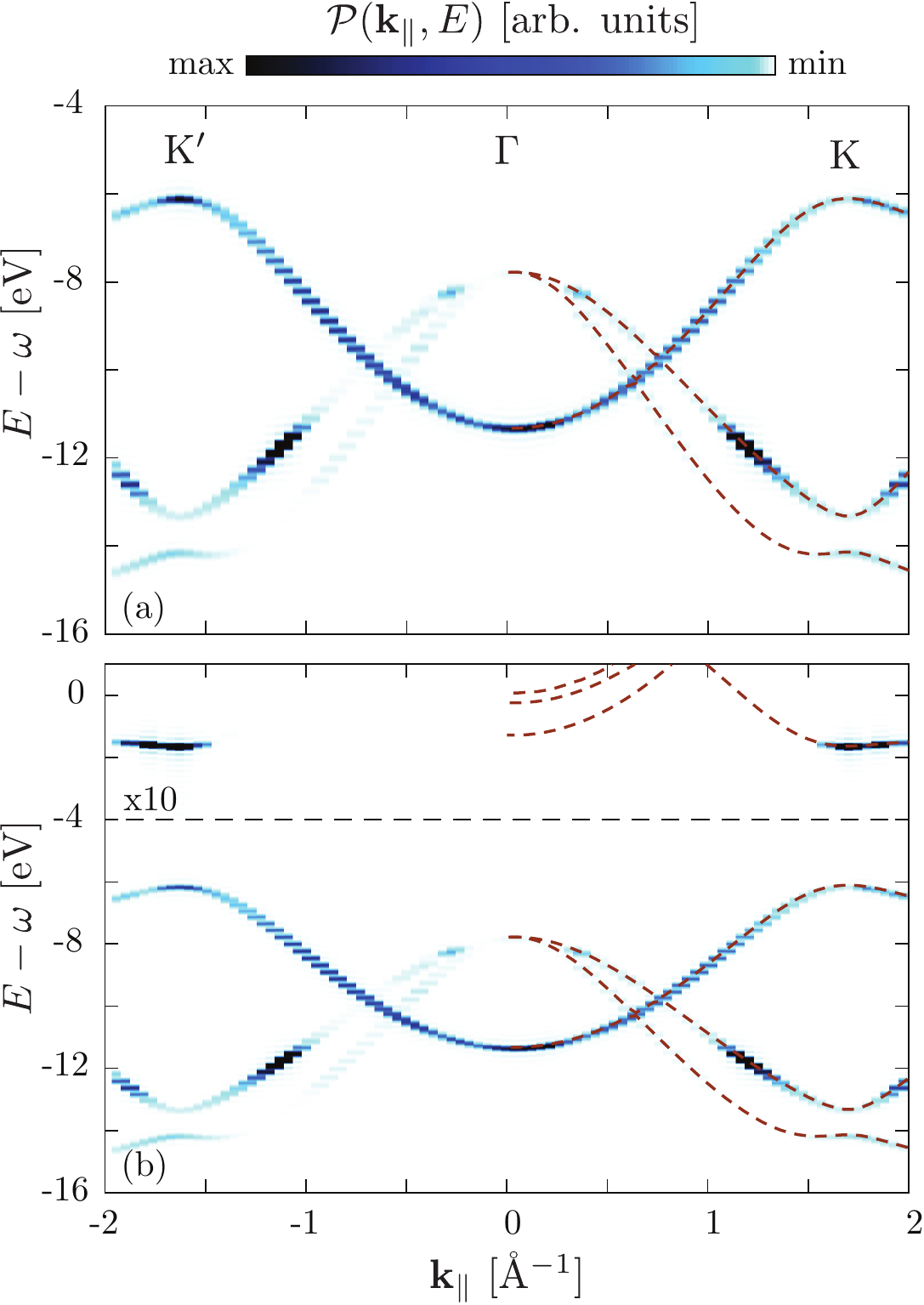}
  \caption{\label{fig:hbn}
    Pump-probe ARPES on monolayer hBN. 
    In (a) ARPES for the system in equilibrium probed by a 24~fs pulse 
    polarized at 45$^\circ$ from the surface plane with 
    $\boldsymbol{\epsilon}_\parallel=\mathbf{a}_1$, 
    $\omega=40.8$~eV, and $I=10^{10}$~W/cm$^2$.
    In (b) we show tARPES for the system pumped with a 20~fs pulse polarized in the plane of the 
    surface $\boldsymbol{\epsilon}=\mathbf{a}_1$ 
    resonant with the gap at K, $\omega=4.46$~eV, $I=2\times10^{11}$~W/cm$^2$ and 
    probed by the same laser of panel (a) right after the pump is switched off, $\Delta t= 20$~fs.
    ARPES signal on the conduction band is magnified by a factor 10.
    The band structure is overlaid in red on both panels.
    }
\end{figure}

First we probe the system in its ground state with a 24~fs UV pulse linearly polarized 45$^\circ$ 
off-plane along one of the lattice vectors $\boldsymbol{\epsilon}_\parallel=\mathbf{a}_1$ with 
$\omega=40.8$~eV, and $I=10^{10}$~W/cm$^2$. The simulations have been carried out with a grid spacing 
0.36~a.u. and a lattice constant of $a=4.76$~a.u..
The resulting ARPES in Fig.~\ref{fig:hbn}~(a) is in good agreement with  
experimental data~\cite{Usachov:2012ft} and with the DFT band structure.

Next, we first pump the system with a laser pulse and probe it right after the pump is switched off.
To this end we used a 20~fs in-plane pump pulse $\boldsymbol{\epsilon}=\mathbf{a}_1$ with  $\omega=4.46$~eV, $I=2\times10^{11}$~W/cm$^2$ and then probe with a laser 
delayed of $\Delta t = 20$~fs.
The pump pulse is resonant with the gap at K and therefore it can excite electrons from 
the valence to the conduction band. 
This excitation is confirmed by the tARPES spectrum in Fig.~\ref{fig:hbn}~(b) where we observe 
a signal from electrons located on the conduction bands around both K and K$^\prime$.
This is a clear indication of a resonant population transfer from valence to conduction band 
and a simple demonstration of how the method presented here can be used to simulate the full 
dynamics of a pump-probe ARPES experiment.

\section{Conclusions} 
\label{sec:conclusions}

In this paper we have presented the t-SURFFP method -- a novel ab-initio technique to simulate spin and time-resolved ARPES on semi-periodic systems based on TDDFT. This method makes no assumption on the probe pulse leaving polarization, energy and pulse shape (envelope) free to be chosen to closely match experimental conditions. 
The ionization dynamics is fully simulated by time-propagation of the electronic density under presence of the classical probe field fully accounting for electron-electron scattering, electron-ions scattering, surface image charge effects, classical screening and other dynamical effects, thus naturally including mean free path and matrix element effects. 
Quantum mechanical exchange and correlation effects, however, are approximated within the TDDFT framework through the exchange and correlation density functional. 
This latter approximation imposes some limitations to systematically describe strongly correlated systems and many-body effects, but for some cases specialized functionals exists. By accounting for the spin degree of freedom of the electrons this method also utilizes the extension of TDDFT to spin density dynamics to simulate spin-resolved ARPES measurements.
The fully flexible definition of the external fields together with the first principles propagation of the density allows to create complex pump-probe setups where the electronic structure is excited separately by arbitrary pump pulses. This feature in particular allows for the first time the fully ab initio study of non-equilibrium electron dynamics under pump probe conditions with a large variety of applications.

We have thus introduced a versatile and general computational method for the ab-initio study and simulation of photoemission experiments.


\section{Aknowledgments} 
\label{sec:aknowledgments}

We acknowledge financial support from the European Research Council (ERC-2015-AdG-694097), Spanish grant 
(FIS2013-46159-C3-1-P), Grupos Consolidados (IT578-13), AFOSR Grant No. FA2386-15-1-0006 AOARD 144088, 
and European Union’s Horizon 2020 Research and Innovation program under Grant Agreements no. 676580 
(NOMAD) and 646259 (MOSTOPHOS). H.H. acknowledges support from the People Programme (Marie Curie Actions) of the European Union's Seventh Framework Programme FP7- PEOPLE-2013-IEF project No. 622934.


\bibliography{paper.bib}

\end{document}